# Enhancement of 3$^{rd}$-harmonics generation during ultrashort pulse diffraction in multi-layer volume-grating

**Tan Yizhou,** [1,*]   **Yang Yisheng,**[2]   **Liao Jiali,**[2]  **and Zhang Ming**[1]

[1] *College of Mechatronic Engineering and Automation, National University of Defense Technology, 410073, China*

[2] *College of Science, National University of Defense Technology, Changsha, 410073, China*

*tanyizhou@126.com*

**Abstract:** Successful phase-matching methods for Third Harmonics Generation (THG) include phase-matching in birefringent crystal and quasi-phase-matching (QPM) in crystal with periodically poled domains. However, these methods are not feasible in some isotropic materials (e.g. fused silica and photosensitive silicate glass). It was known that volume-grating in isotropic materials can independently generate frequency-converted waves. One of disadvantages of single-layer volume-grating is that the brightness of harmonic emission can not be enhanced by increasing the grating thickness. In this paper, a THG device with stratified sub-gratings was designed to enhance THG in isotropic materials: several sub-gratings were arranged parallel, and the grating-figures misalignment between neighboring sub-gratings was pre-fabricated. In terms of extension of interaction length in THG, our multi-layer sub-grating is formally equivalent to the multi-layer periodically poled crystal (e.g. lithium niobate) in conventional QPM approach. According to the calculation results, the *N*-layer ($N > 2$) can, in principle, generate TH output intensity of $N^2$ times stronger than single-layer volume-grating does, also compared to *N* times stronger than *N*-layer without figures-misalignment. The effect of random fabrication error in grating thickness on normalized conversion efficiency was discussed.

Harmonic generation and mixing, Pulse shaping, Optical devices: Nonlinear optical devices, Holographic optical elements, Volume gratings

## 1. Introduction

The direct third-harmonic-generation ($\omega + \omega + \omega \rightarrow 3\omega$) is becoming attractive to the practical applications, such as direct determination of the $\chi^{(3)}_{bulk}$ components of centrosymmetric media (such as silica and glasses) [1], scanning laser third-harmonic-generation (THG) microscopy and so on.

Some isotropic materials (e.g. fused silica and glasses) have low nonlinearity and larger material dispersion, resulting in lower signal strengths in direct THG process. Bulk-enhanced THG was first observed in isotropy media containing diffractive grating [2, 3]. Soon afterwards, V. I. Smirnov and L.N. Glebova et al [4] found that an efficient THG process occurred in grating pre-recorded in photosensitive silicate glass of 1~2 mm thick, if Bragg diffraction condition of volume grating was satisfied. Since then, bulk-enhancement of THG with refractive-index-grating has been studied in detail, and experimental results were repeatable [4-6].

One of disadvantages of single-layer grating used in [4-6] is that the brightness of harmonic emission can not be enhanced by increasing the grating thickness. In this paper, we design a THG device with stratified sub-gratings to overcome the above shortcoming. As shown in Fig.1a, we arrange *N*-layers holographic sub-gratings ($N \geq 2$) into a longitudinal periodic structure with dislocation-interfaces. In terms of extension of interaction length in THG, our multi-layer sub-grating is formally equivalent to the multi-layer periodically poled crystal (e.g. lithium niobate) in conventional quasi-phase-matching (QPM) approach [7].

## 2. Optically induced QPM in multi-layer grating

Two pulse-shaping processes occur by turns in multi-layer grating as follows:

(*i*) When pump waves enter each layered sub-grating, diffracted waves with transmitted waves appear. They overlap to produce double-beam interference patterns, which create a driving-field with arc-shape (or sawtooth-like function) phase-modulation profile.

(*ii*) There is misalignment between neighbouring sub-gratings in longitudinal periodic structure. When pump-pulse is traveling across interfaces with such non-symmetry, a step-function phase-shift (hereafter called "phase-jump") arises transiently.

Consequently, a pulse with Gaussian waveform is reshaped in new one with "arc-shape modulation + phase-jump" profile. QPM is achieved as follows: in each sub-grating, both pump-wave and 3$^{rd}$-harmonic wave are spatially modulated, so the phase slips between them are partially corrected. In addition, the relative phase of harmonics generated in neighbouring sub-gratings is periodically corrected by the phase-jumps; hence in-phase emission from many individual layers can be coherently summed.

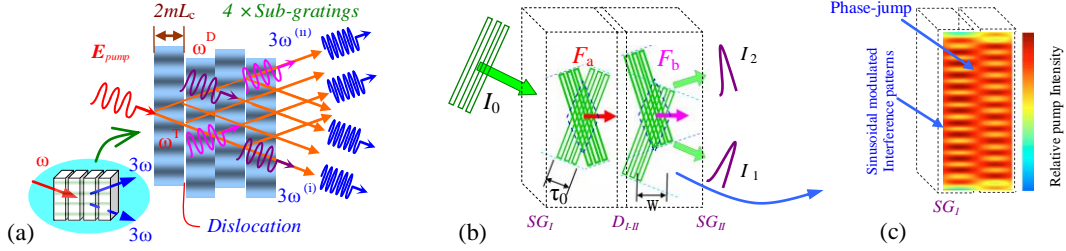

Fig. 1 (Color online) Third-harmonic generation in multi-layer grating with dislocation

(a) Structure diagram of multi-layer grating: sub-gratings are arranged parallel. The dislocation was pre-fabricated between neighboring sub-gratings. The $E_{pump}$ is the pump beam, $3\omega^{(i)}$ and $3\omega^{(ii)}$ are generated harmonics waves;

(b) Schematic drawing of pulse-shaping: (i) in single-layer grating, the transmitted beam $I_1$ and diffracted beam $I_2$ overlap and generate a set of double-beam interference patterns $F_a$ and $F_b$ in sub-gratings; (ii) when ultrashort laser pulse is traveling with velocity $V_i = c/(n_\omega \cdot \cos\alpha_{cB})$ across the interface $D_{I-II}$, the dislocation causes a lateral shift between the patterns $F_a$ and $F_b$, which corresponds to a phase-jump in time-domain;

(c) Simulation result illustrates the dynamic process (Media 1) — a pulse is travelling through multi-layer grating, and the driving field distributions are laterally shifting at each interface.

## 3. Growth of the harmonic signal in single-layer grating

As the basic element of multi-layer grating, each sub-grating can independently generate frequency-converted waves. In the experimental setup (Fig. 2), a refractive-index-grating (represented by Eq. (a1)) was pre-fabricated in photo-thermo-refractive glass of thickness in 1~2mm. This grating splits the fundamental wave $E_{pump}$ into transmitted beam $E_1^{(T)}$ and diffracted beam $E_2^{(D)}$ in the y-z plane. The generated non-collinear harmonic waves $3\omega^{(a)}$ and $3\omega^{(b)}$, together with its diffraction components $3\omega^{(c)}$ and $3\omega^{(d)}$, were observed on screen. THG experiments at an incident intensity of $5\times10^{11}$ ~$10^{12}$ W/cm$^2$ (120~150 *fs*) have been verified at different wavelengths, such as 775nm→258nm, 800nm→267nm, 1550nm →517nm and so on [4-6].

Using grating as pulse-shaper, QPM in THG process is optically induced as follows:

In Fig. 1b and Fig 2, the interference field consists of transmitted wave $\vec{E}_1(y,z) = E_1 \exp[j(\vec{k}_{1y} \cdot y + \vec{k}_{1z} \cdot z)] \cdot \exp[j(\omega t)]$ and diffracted wave $\vec{E}_2(y,z) = E_2 \exp[j(-\vec{k}_{2y} \cdot y + \vec{k}_{2z} \cdot z)] \cdot \exp[j(\omega t)]$.

They overlap with each other. Neglecting absorption, the modulated amplitude and shifted phase of pump-field are:

$$E_{dri}(y',z') = \{E_1^2[1+(E_2/E_1)^2+2(E_2/E_1)\cos\rho]\}^{1/2}, \qquad (1)$$

$$\phi_{dri}(y',z') = \tan^{-1}\{[(E_2/E_1)\sin\rho]/[1+(E_2/E_1)\cos\rho]\} \qquad (2)$$

The phase-shift between the ω and 3ω waves consist of two terms: the first term $\Delta\Phi_0(z')$ is the linear growth in the phase change due to material dispersion, the second term $\Delta\phi_{dri}(z')$ describes the optically induced phase-modulation. Following the theory of grating-assisted-phase-matching [7, 10], the slowly-varying envelope approximation of the 3$^{rd}$-harmonic field after propagation length *L* in medium is given by

$$E_{3\omega}(z') \approx \Gamma \cdot \int_0^L (A_\omega)^3 \cdot \chi_{eff}^{(3)} \cdot \exp[-j(\Delta\Phi_0 - \Delta\phi_{dri})] \cdot dz' \qquad (3)$$

where $\Gamma = j3\omega/2cn_3$, assuming nonlinear susceptibility $\chi_{eff}^{(3)}$ remains unchanged in medium with refractive-indix-grating. If phase-shift $\Delta\phi_{dri}(z')$ has a modulation period that corresponds to two coherence lengths $2L_c$ and a slope that corresponds to $-\Delta\Phi_0(z')$, the harmonic emissions in Eq.(3) will be maximized, leading to a linear growth of the harmonic signal.

Considering single-layer grating has reached physical limit in performance improvements, in next section we utilize multi-layer gratings to enhance the brightness of harmonic emission.

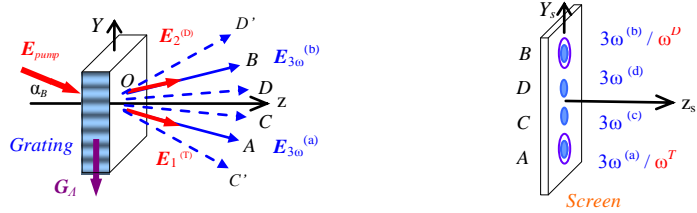

Fig.2. (Color online) Non-collinear Third-Harmonic Generation in single-layer sub-grating

Schematic diagram of experimental setup [6-8]: $E_1^{(T)}$, $E_2^{(D)}$ are transmitted and diffracted fundamental waves; $E_{3\omega}^{(a)}$ and $E_{3\omega}^{(b)}$ are generated harmonic waves. The original TH waves $3\omega^{(a)}$ and $3\omega^{(b)}$ emerge along directions $OA$, $OB$, and its diffraction components (e.g. $\sigma_0$, $\sigma_{+1}$, $\sigma_{+2}$ and so on) are spread in $OC$, $OC'$ and $OD$, $OD'$ respectively.

## 4. Coherent buildup of the harmonic signal in multi-layer grating

The design theory of "stratified volume diffractive optical element" was proposed by D. M. Chambers et al [9, 10]. As an application example, beam-scanning elements for space-based lidar had been studied, which was made up of multi-layer binary gratings interleaved with homogeneous intermediate layers.

The multi-layer grating in Fig. 3a looks similar in shape to the previous device reported in [9, 10], but our device is characterized in misalignment of neighbouring sub-gratings. Such unsymmetrical structure is called "grating-figure dislocation", because it is formally equivalent to local dislocation in crystal.

*4.1 Phase-jumping at interface with dislocation*

In sub-grating in Fig.3b, the phase of driving pulse is described by Eq. (2). Subsection $SG_{II}$ is laterally shifted relative to subsection $SG_I$. Here the geometric displacement $\Lambda/4$ (in $OY$ orientation) corresponds to spatial phase-shift $\pi/2$ of the grating-fringes. When laser pulse with velocity $V_i = c/(n_\omega \cdot \cos\alpha_{cB})$ is traveling across the interface $D_{I-II}$, the grating-figures dislocation in spatial domain causes a phase-jumping $\Delta\delta_{jum}$ in temporal domain accordingly.

We use unit step-function $U(t-t_0)$ to turn on the phase-jump at each interface; assuming $m^{th}$ interface $D_{I-II}$ was located at $z = m(\Delta d + \Delta d')$, the phase function was given by

$$\Delta\phi_{jum}(z,t) = U(t-t_m) \cdot \Delta\sigma_{jum} = \Delta\sigma_{jum}, \quad \text{if } t = t_m; \quad \text{Jump-position in time: } t_m = (m \cdot \Delta d/\cos\alpha_B) \cdot (n/c)$$
$$U(t-t_m) \cdot \Delta\sigma_{jum} = 0, \quad \text{if } t \neq t_m, \qquad (4)$$

where $t$ is pulse flight-time; $t_m$ is position of $m^{th}$ interface in time coordinates; $(c/n)$ is wave velocity in medium. $\Delta d$ is the efficient thickness of sun-grating, $\Delta d'$ is thickness of intermediate layer (refractive-index-matching-material) between two sub-gratings, $\Delta d' \ll 1$.

The $\Delta\sigma_{jump} = \sigma^{(n+1)}(y,z_{n+1}) - \sigma^{(n)}(y,z_n)$ denotes the phase-difference of laser pulse between the left side and right side of the interface $D_{I-II}$, which is calculated by

$$\sigma_{jum}^{(n)}(z_s) = \tan^{-1}[(A_1 \sin\sigma_1 + A_2 \sin\sigma_2)/(A_1 \cos\sigma_1 + A_2 \cos\sigma_2)] \qquad (5)$$

where $A_i$ and $\sigma_i$ denote the amplitude and phase of foundational waves in sub-grating $SG_j$ ($i = 1, 2, j= I, II$). The details are in Appendix.

*4.2 Optically induced QPM in multi-layer grating*

In THG device made up of multi-layere gratings, the total driving-field may be rewritten in combinations of Eqs.(1), (2) and (4) as $\tilde{E}_{Tot} = A_\omega \exp j[k'z' - \omega t + \phi(z)] \exp j[\phi_{dri}(y,z)] \exp j[\Delta\phi_{jum}(z^s)]$.

Rewriting Eq. (3), the 3$^{rd}$-harmonic field in stratified gratings is given by

$$E_{3\omega}(z') \approx \Gamma \cdot \int_0^L (A_\omega)^3 \cdot \chi_{eff}^{(3)} \cdot \exp[-j(\Delta\Phi_0 - \Delta\phi_{dri} - \Delta\phi_{jump})] \cdot dz' \qquad (6)$$

There are two types of phase-mismatch sources in medium. As the main source, phase-mismatch $\Delta\Phi_0$ results naturally from different phase-velocities of $3\omega$ and $\omega$ waves, $\Delta\Phi_0 = \Delta k' \cdot z_c'$ and $\Delta k' = 3k_\omega - k_{3\omega} = \pi/L_c$. The second type can be controlled artificially, which was described by Eq. (2) and (4): $\Delta\phi_{dri}$ — arc-shape (or sawtooth-like function) phase-modulation caused by diffraction in layered sub-grating, $\Delta\phi_{jum}$ — phase-jump introduced by dislocation between sub-gratings.

An inspiring fact is that multi-layer grating has capability to balance the natural phase-mismatch with the artificial phase-shift. When arc-shape (or sawtooth-like function) modulation and phase-jump combined together, two kinds of artificial phase-shift mechanism (related to $\Delta\phi_{dri}$ and $\Delta\phi_{jum}$) do not influence each other, because they are staggered in time-domain.

Both $\Delta\phi_{dri}$ and $\Delta\phi_{jup}$ can be separately adjusted by changing the geometric parameters (such as the incident angle or thickness of sub-grating), so that magnitude of ($\Delta\phi_{dri} + \Delta\phi_{jum}$) is close to net mismatching $\Delta\Phi_0$. And the Eq. (6) can be integrated:

$$E_{3\omega}(L) \propto j\Gamma(A_\omega)^3 \chi_{eff}^{(3)} \exp[-j\frac{L}{2}(\Delta k' - k_{dri} - k_{jum})] \cdot \mathrm{sinc}[\frac{L}{2}(\Delta k' - k_{dri} - k_{jum})] \qquad (7)$$

where $\mathrm{sinc}(x) = \sin(x)/x$; ($k_{dri} + k_{jum}$) corresponds to the combined phase-shift ($\Delta\phi_{dri} + \Delta\phi_{jum}$).

In practical terms, the refractive-indices $n_{(800\ nm)} \approx 1.49$ and $n_{(267\ nm)} \approx 1.54$ in silicate glass, so coherence-length $L_c = \lambda/[6(n_{3\omega} - n_\omega)] \approx 2.7$ μm in frequency conversion 810 nm → 270 nm. Two sets of experimental parameters were used in Ref. [4-6] to satisfy the QPM requirement $\Lambda_{patt} = 2mL_c$ in individual sub-grating: such as first-order QPM ($m = 1$, grating period $\Lambda_g \approx 0.54$ μm, Bragg angle $\alpha_B \approx 15.88°$) and third-order QPM ($m = 3$, $\Lambda_g \approx 0.81$ μm, $\alpha_B \approx 12.92°$). Their conversion-efficiency plots are shown in Fig. 3c.

Phase-matching for dislocation-interface (arranged in longitudinal direction, i.e. oz axis) must be compatible with phase-matching for lateral sub-gratings, so the QPM condition of dislocation-interfaces is given by $T_{eff} = q(2b+1)\Lambda_{patt}$ and $(2b+1) \cdot \Lambda_{patt} = 2mL_c$; where $T_{eff}$ is effective thickness of sub-grating and $q$, $b$ are positive integer. As a note, both dislocation-interfaces ($\Delta\phi_{jup}$) assisted QPM and individual sub-grating ($\Delta\phi_{dri}$) assisted QPM have same mathematical expressions; the only difference is $T_{eff} \gg \Lambda_{patt}$ (or $q \gg 1$). It means that the interval between neighbouring dislocations is larger, in other words, sub-grating is thicker. The reason is that volume grating (of thickness ~1000μm with diffraction efficiency > 90%) is favourable to achieve QPM in this paper.

In order to sum the TH intensities arising from *N*-layer sub-gratings, without loss of generality, we consider each single-layer as a light-emitting-surface (a thin nonlinear oscillator at the harmonic frequency: $\tilde{E}_m = E_m \cdot \exp[j(3\omega t + \varphi_n)]$). The phase $\varphi_{n+1}$ of the $(n+1)^{th}$ emitting-surface linked in phase with the $n^{th}$ emitting-surface by the phase-jump, which appeared at incident side of each sub-grating. The total amplitude $\tilde{E}_T$ due to *N* such contributions is

$$\tilde{E}_T = E_m \cdot \exp(j3\omega t)\sum_{n=1}^N \exp(j\varphi_n) . \qquad (8)$$

Using theoretical treatment similar to [7, 8], we take the average valve of total intensity $\langle \tilde{E}_T^2 \rangle$:

(*i*) In ideal multi-layer structure, all surface-oscillators have identical phase, the coherent-superposition gives the harmonic amplitude as $\langle \tilde{E}_T^2 \rangle = N^2 \cdot E_\omega^2$. It means that total output TH intensity can, in principle, be increased by a factor of *N* above that for individual single-layer;

(*ii*) In practice, the existence of fabrication error should not be neglected. Foe example, thickness tolerance of individual sub-grating caused slightly irregular interval between two neighboring phase-jumps. *N* waves with random phases are superimposed, leading to the total harmonic signal reduced to $\langle \tilde{E}_T^2 \rangle = N \cdot E_\omega^2$ (where both absorption loss and diffraction loss were ignored for clarity) [7, 8].

The statistical variations in the effective thickness of individual sub-grating are known as longitudinal-period-error $\sigma_T^2$. The plots in Fig. 3d indicate that the conversion-efficiency $\langle \bar{\eta} \rangle$ from $N$-layers device will be reduced by 50%, when longitudinal-period-error of stratified gratings $\sigma_T \geq (1/N^{-2}) \cdot (L_c / \pi)$. For a multi-layer device 9-mm long with thickness of ~1000 μm, $N = 9$, the tolerance should not exceed $0.12\, L_c$, making the micro-fabrication much more difficult.

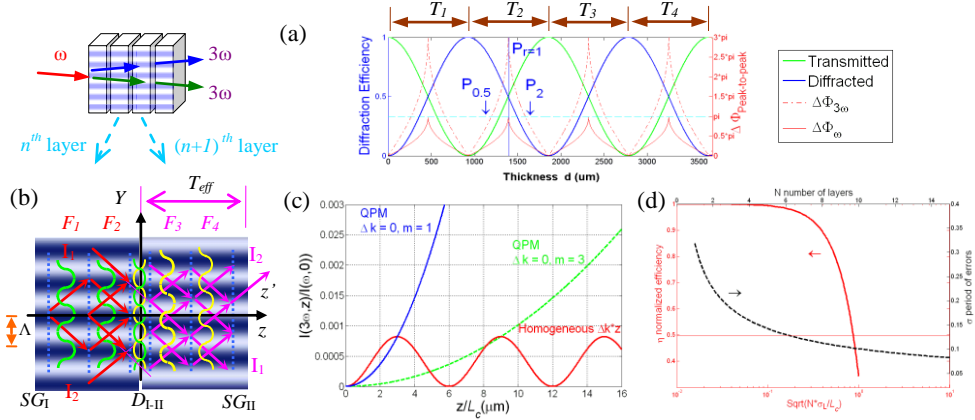

Fig. 3 (Color online) Optically induced QPM in multi-layer grating with dislocation

(a) Configurations of THG device: multiple sub-gratings with thickness $T_i$ are arranged in parallel. Sub-gratings have same spatial period $\Lambda$, so the field distribution in each layer is similar to that in Eq.(1), (2) (Neglecting absorption and scanning).
(b) Schematic diagram of phase-jump between two sub-gratings: laser pulse is traveling across the interface $D_{I\text{-}II}$, the transmitted and diffracted sub-waves from the first subsection create a double-interference pattern in the second subsection. Similar situations occur with the third and fourth subsections in sequence. The induced-grating $F_1$ and $F_2$ (or $F_3$ and $F_4$) in same subsection are keeping in-phase; however a phase-jump raised between the induced-grating $F_2$ and $F_3$ due to the misalignment of adjacent sub-gratings $G_I$ and $G_{II}$.
(c) Relative TH intensity under three different phase-matching conditions: first QPM ($m = 1$), three order QPM ($m = 3$) and phase-mismatching $\Delta k' \cdot z_c' \approx 5$ cm$^{-1}$.
(d) The standard deviation of the longitudinal period (i.e. thickness of sub-grating) for > 50% conversion is plotted (black line), which changes over $N$ in $N$-layers stack.
  As reference, the effect of random imperfections in grating thickness on normalized conversion efficiency is also plotted (red line), which changes over $(N\sigma_T / L_c)^{1/2}$, where the conversion efficiency $\hat{\eta}$ of an imperfect device is normalized to the "ideal" quasi-phase matched device. Assuming the reduction of conversion efficiency duo to longitudinal-period-error is small, the normalized conversion efficiency $\langle \bar{\eta} \rangle$ of a stratified device is given as [7]

$$\langle \bar{\eta} \rangle \approx 1 - (\pi^2 / 15) \cdot N \cdot (\sigma_T^2 / L_c^2),$$ (the angle brackets indicate an assemble average).

## 5. Summary

The THG device consists of multi-layer sub-gratings. An all-diffractive pulse-shaping method in the femtosecond regime is exploited to enhance the generated harmonics: arc-shape (or sawtooth-like function) phase-modulation is produced by single-layer diffractive grating, and phase-jumps arise at each dislocation-interface between neighbouring sub-gratings.

It is to our knowledge the first QPM scheme that phase-mismatching is corrected by transverse periodic structures ($N$-pieces sub-gratings) and longitudinal periodic structure (equally spaced interfaces with dislocation) alternately; and the grating-fringes dislocation plays a key role in keeping $N$-harmonic oscillators in-phase over a longer interaction length.

Most of all-optical QPM approaches require external devices to precisely control or adjust the modulation-profile of pump-pulses [7, 8]. The key advantage of multi-layer grating is its simplicity in pulse-shaping, since favorable modulation-profile is produced through direct space-to-time transformation in the gratings (rather than by external pulse-shapers).

The conventional approaches (such as birefringence-phase-matching and quasi-phase-matching in crystal with periodically poled domains) are not feasible for isotropic materials used in this work. Beyond fundamental research, optically induced QPM in this paper could be a relatively easy, cheap solution to the phase-matching puzzle in bulk-isotropic materials, despite it is less efficient than the "truly" phase-matched process.

**Acknowledgments**

This work was part funded by the National Natural Science Foundation of China (Grant No. 61273202) and Project of the 3rd Innovation Experiment & Inquiry Learning Program (No.2010-5, Education Department of Hunan, China).

**Appendix**

Assuming $n_{ave}$ is the average value of refractive-index, $\Delta n_2$ is its modulation-amplitude; $\Lambda$ is spatial period of grating. $\Theta(y,z)$ is spatial phase of grating-figures. Fixed-grating in Fig.1b and Fig.3b is approximately represented by

$$n_{fix}(y,z) \approx n_{ave} + \Delta n_2 \cdot \cos[(2\pi y/\Lambda + \Theta_{fix}^{(m)}(y,z)],  \qquad (a1)$$

Using finite-element method [11], a thick refractive-index-grating was partitioned into subsections (also called slabs) of thickness $\Delta d$. The magnitude of $\Delta d$ was sufficiently small, so that each slab acts as a thin grating. As shown in Fig. 1b and Fig. 3b, a laser pulse is traveling from $m^{th}$ slab to neighbouring slab, diffracted wave $I_2(y, z)$ and transmitted wave $I_1(y, z)$ appeared in $(m+1)^{th}$ slab. Two waves overlap to form interference patterns, which induce a transient grating in medium. This induced grating is approximately represented by

$$n_{ind}(y,z) \approx n_{ave} + n_{kerr} \cdot \cos[2\pi y/\Lambda + \Xi_{ind}^{(m)}(y,z)], \qquad (a2)$$

where $n_{kerr}$ is the Kerr coefficient of medium. $\Xi_{ind}^{(m)}(y,z)$ is spatial phase of grating-figures.

We follow the theory of two-wave mixing [12-14] to calculate the intensity of fundamental waves $I_1(y,z)$ and $I_2(y,z)$. The solutions to Kogelnik coupling wave equation in grating are:

$$I_1(y,z) = I_1 \cos^2(\varsigma z) + I_2 \sin^2(\varsigma z) - (I_1 I_2)^{1/2} \sin(2\varsigma z) \sin(\Delta\Omega_{spa})$$
$$I_2(t,z) = I_2 \cos^2(\varsigma z) + I_1 \sin^2(\varsigma z) + (I_1 I_2)^{1/2} \sin(2\varsigma z) \sin(\Delta\Omega_{spa}) \qquad (a3)$$

where the $\zeta$ is the coupling constant ($\zeta = \pi\Delta\varepsilon/2(\varepsilon)^{1/2}\lambda\cos\theta$). The first two terms describe diffracted and transmitted waves. The third terms describe transient energy-transfer between two beams, which are related to the spatial phase-difference $\Delta\Omega_{spa}^{(m)} = \Theta_{fix}^{(m+1)} - \Xi_{ind}^{(m)}$ [14].

<u>Case I</u>: grating-fringes of neighbouring slabs are aligned to each other $\Theta_{fix}^{(m)} = \Theta_{fix}^{(m+1)}$:

According to [13-15], wavefront-adaptation of induced-fields is a intrinsic property of diffraction effect, which leads to the induced-grating being always coinciding with the fixed-grating-fringes in $m^{th}$ slab (i.e., $\Xi_{idu}^{(m)} \approx \Xi_{idu}^{(m+1)}$ and $\Delta\Omega_{spa} \approx 0$ for neighbouring slabs in ordinary volume grating). Bragg scattering occurs, the pump-field is described by Eqs.(1) and (2).

<u>Case II</u>: there are misaligned grating-fringes between neighbouring slabs, $\Theta_{fix}^{(m)} \neq \Theta_{fix}^{(m+1)}$:

Two-wave mixing experiment results and theory indicate that the presence of non-zero spatial phase-difference $\Delta\Omega_{spa}$ leads to nonreciprocal transfer of energy between the beams. More importantly, spatial phase-difference $\Delta\Omega_{spa}$ can be introduced to Kerr media by use of external conditions (e. g. using the Lorentz force to move free-carrier grating in a medium, or tuning the frequency of laser beams to make the double-beam interference pattern moving; both approaches are referred to as "moving grating technologies" ) [13-15].

In this work, an alternative approach to form spatial phase-shift of grating is proposed: the key idea is that pre-fabricated misalignment of neighbouring fixed-gratings is equivalent to moving a grating in time-domain [13]. In experiment in Fig.2, the pulse laser duration $\tau_0$ is ~100 $fs$, and the electronic response of Kerr media is less than ~1 $fs$. Therefore, an induced-grating is a sheet in thickness of ~20μm, and is traveling with velocity $V_i = c/(n_\omega \cdot \cos\alpha_{cB})$. So we can utilize geometric dislocation to implement direct space-to-time conversion for pulse-shaping.

At interface $D_{I-II}$, we denote spatial phase-difference by $\Delta\Omega_{spa}^{(\theta)} = \Theta_{fix}^{(n+1)} - \Theta_{ind}^{(n)}$ to distinguish from the spatial phase related to $\Xi_{ind}^{(m)}(y,z)$ in Eq. (a2). As shown in Fig. 3b, two neighbouring slabs located in same sub-grating $SG_I$, the induced-grating sheets $F_1$ and $F_2$ are in phase; however, the induced-grating sheets $F_2$ and $F_3$ are out of phases ($\Xi_{idu}^{(m)} \neq \Xi_{idu}^{(m+1)}$) owing to the misalignment of sub-grating $SG_I$ and $SG_{II}$, and phase-jump happens accordingly. For example, if we shift laterally sub-grating $SG_{II}$ to $SG_I$ with displacement of $\Lambda/4$ of grating-fringes, phase-deference $\Delta\Omega_{spa}^{(\theta)} = \pi/2$ will be introduced between $n^{th}$-layer and $(n+1)^{th}$-layer sun-gratings. According to Eq. (a3), in theory, such a dislocation will cause the maximum energy transfer between two beams (Neglecting absorption and scanning).

For clarity of the description, two waves in Eq. (a3) are re-written in complex notation as

$$E_1(y,z) = A_1(y,z) \cdot \exp[-j(\omega t + \sigma_1)] + c.c.$$
$$E_2(y,z) = A_2(y,z) \cdot \exp[-j(\omega t + \sigma_2)] + c.c.$$
(a4)

where $A_1$ and $A_2$ denote the amplitude, $\sigma_1$ and $\sigma_2$ denote the phase respectively. Ref. [12, 13, 15] provided more detailed mathematical expressions of $\delta_i$ and $A_i$ ($i=1, 2$).

The total electric field strength of two-wave superposition is $\vec{E}_{jum}(y,z) = \vec{E}_1(y,z) + \vec{E}_2(y,z)$
$= \text{Re}\{A_{jum}(y,z) \cdot \exp[-j(\omega t + \sigma_{jum})]\}$. And the amplitude and phase of $\vec{E}_{jum}(y,z)$ in $m^{th}$ slab are

$$A_{Tot}^2(y,z) = A_1^2(y,z) + A_2^2(y,z) + 2A_1 A_2(y,z) \cdot \cos(\sigma_1 + \sigma_2),$$
(a5)
$$\sigma_{jum}^{(n)}(z_s) = \tan^{-1}[(A_1 \sin\sigma_1 + A_2 \sin\sigma_2) / (A_1 \cos\sigma_1 + A_2 \cos\sigma_2)]$$
(a6)

In a word, the spatial phase-shift $\Delta\Theta_{fix}$ (pre-fabricated between $(N+1)^{th}$ and $N^{th}$ sub-gratings) has function to introduce temporal phase-shift to the pulse, i.e. $\Delta\sigma_{jump} = \sigma^{(n+1)}(y,z_{n+1}) - \sigma^{(n)}(y,z_n) \neq 0$.

Due to limited space, this paper only analyze phase-change in driving pulse, and not enter the discussion of the pulse broadening and waveform distortion for the moment. As a note, theoretical results in Eq. (7-8)) and (a3-a6) still hold true by taking into account the wavefront correction, because spatial-temporal distortions have limited impact on periodically re-setting of the nonlinear interaction phase in THG process.